\documentclass[useAMS,usenatbib]{mn2e}

\usepackage{dsfont}
\usepackage{amsmath}
\usepackage{amssymb}
\usepackage{graphicx}
\usepackage{verbatim}
\usepackage{natbib}
\usepackage{eucal}
\usepackage{calligra}
\usepackage{subfigure}

\usepackage[usenames,dvipsnames]{color}

\usepackage{amsfonts}
\usepackage{color}
\usepackage[normalem]{ulem}
\usepackage[T1]{fontenc}

\DeclareMathAlphabet{\mathscr}{OT1}{pzc}%
                                 {m}{it}
                                 
\newcommand{\mnras}{MNRAS}
\newcommand{\jcap}{JCAP}
\newcommand{\apj}{ApJ}

\newcommand{\aap}{A\&A}

\newcommand{\nat}{Nature (London)}

\newcommand{\be}{\begin{equation}}
\newcommand{\ee}{\end{equation}}
\newcommand{\bes}{\begin{equation*}}
\newcommand{\ees}{\end{equation*}}
\newcommand{\bea}{\begin{eqnarray}}
\newcommand{\eea}{\end{eqnarray}}
\newcommand{\beas}{\begin{eqnarray*}}
\newcommand{\eeas}{\end{eqnarray*}}

\newcommand{\de}{{\rm d}}
\newcommand{\mpch}{\;{\rm Mpc}/h}

\newcommand{\rhosub}{\rho_{\rm subhalo}}
\newcommand{\rhohost}{\rho_{\rm host}}
\newcommand{\rtidal}{r_{\rm tidal}}
\newcommand{\tdyn}{t_{\rm dyn}}
\newcommand{\tquench}{t_{\rm quench}}
\newcommand{\tinfall}{t_{\rm infall}}

\def\redmagic{redMaGiC}
\def\redmapper{redMaPPer}

\begin{document}

\title[Tidal Stripping in redMaPPer clusters]
{Tidal stripping as a test of satellite quenching in redMaPPer clusters}
\author[Fang et al.]
{\parbox{\textwidth}{
  Yuedong~Fang$^{1,}$\thanks{E-Mail: yuedong@sas.upenn.edu},
  Joseph~Clampitt$^{1}$,
  Neal~Dalal$^{2}$,
  Bhuvnesh~Jain$^{1}$,
  Eduardo~Rozo$^{3}$,
  John~Moustakas$^{4}$,
  Eli~Rykoff$^{5}$} \\
  \vspace{0.2cm}\\
  $^1$Department of Physics and Astronomy, Center for Particle Cosmology, \\
  \;University of Pennsylvania,
  209 S. 33rd St., Philadelphia, PA 19104, USA \\
  $^2$Department of Astronomy, University of Illinois at Urbana-Champaign, 1002 West Green St., Urbana, IL 61801 USA \\
  $^3$Department of Physics, University of Arizona, 1118 E 4th St, Tucson, AZ 85721 USA \\
  $^4$Department of Physics and Astronomy, Siena College, Loudonville, NY 12211, USA \\
  $^5$SLAC National Accelerator Laboratory, Menlo Park, CA 94025 USA}

\maketitle

\begin{abstract}
When dark matter halos are accreted by massive host clusters, strong gravitational tidal forces begin stripping mass from the accreted subhalos.
This stripping eventually removes all mass beyond a subhalo's tidal radius, but the unbound mass remains in the vicinity of the satellite for at least a dynamical time $\tdyn$.
The N-body subhalo study of Chamberlain et al. verified this picture and pointed out a useful observational consequence: measurements of subhalo correlations beyond the tidal radius are sensitive to the infall time, $\tinfall$, of the subhalo onto its host.
We perform this cross-correlation measurement using $\sim$ 160,000 red satellite galaxies in SDSS redMaPPer clusters and find evidence that subhalo correlations do persist well beyond the tidal radius, suggesting that many of the observed satellites fell into their current host less than a dynamical time ago, $\tinfall < \tdyn$.
Combined with estimated dynamical times $\tdyn \sim 3-5$ Gyr and SED fitting results for the time at which satellites stopped forming stars, $\tquench \sim 6$ Gyr, we infer that for a significant fraction of the satellites, star formation quenched before those satellites entered their current hosts.
The result holds for red satellites over a large range of cluster-centric distances $0.1 - 0.6 \mpch$.
We discuss the implications of this result for models of galaxy formation. 
\end{abstract}


\date{Accepted . Received ; in original form }


\maketitle

\label{firstpage}

\section{Introduction}

Clusters of galaxies -- composed of tens to hundreds of satellite galaxies orbiting within a dark matter halo -- are the most massive virialized objects in the Universe.
Their average dark matter profiles and halo masses have been measured in great detail (e.g., \citealt{mandelbaum06, johnston07, mandelbaum08, sheldon09, rozo09, mandelbaum10}).
Recent work has measured finer features of the mass distribution within and around cluster halos.
The splashback radius predicted by \citet{diemer14} and \citet{adhikari14} was recently detected by \citet{more16}.
Halo assembly bias \citep{sheth04, gao05, wechsler06, dalal08} has also been detected by the same group \citep{miyatake16, more16}.
Evidence for cluster ellipticity from weak lensing has been measured on small samples of $\sim 20$ clusters \citep{oguri10, donahue16} as well as large samples of several thousand clusters \citep{evans09, clampitt16b}.
Similarly, filaments between individual pairs of massive clusters \citep{dietrich12, jauzac12} and thousands of group and cluster pairs \citep{clampitt16a} have recently been detected using weak gravitational lensing, supplementing similar measurements with the cluster galaxy distribution \citep{zhang13}.

While this work has been accomplished with existing data such as the Sloan Digital Sky Survey\footnote{\texttt{http://www.sdss.org}} (SDSS), the next generation of surveys including the Dark Energy Survey\footnote{\texttt{http://www.darkenergysurvey.org}} (DES; \citealt{melchior15, nord15, rykoff16}), Hyper Suprime-Cam\footnote{\texttt{http://www.naoj.org/Projects/HSC/}} (HSC, \citealt{miyazaki15}) and the Kilo Degree Survey\footnote{\texttt{http://kids.strw.leidenuniv.nl}} (KiDS, \citealt{viola15}) have also begun producing results on galaxy clusters.
Within a few years these new surveys will provide much larger cluster samples than were previously available.

Compared to their massive host clusters, the properties of cluster subhalos are less well-studied.
Subhalos were once isolated dark matter halos, each with their own central galaxy, before gravitational forces pulled them inside much larger neighboring halos.  Many such accreted halos are tidally destroyed and become indistinguishable as distinct entities, but subhalos are the survivors that persist as satellites of the larger cluster halo.
Since the lensing and clustering signals of subhalos are dwarfed by their massive host clusters, it is more difficult to study their detailed mass distributions.
Recent progress in determining weak lensing masses of subhalos has been made by \citet{li15} using CFHT Stripe-82 data \citep{erben13} and by \citet{sifon15} with KiDS data \citep{kuijken15}.
While simulation studies of subhalos predict that tidal forces from the host cluster will strip mass from the subhalo outskirts \citep{hayashi03, gao04}, the recent lensing studies were not able to identify a tidal radius beyond which mass was stripped.
However, the \citet{li15} measurements hinted that subhalos closer to the cluster center -- where tidal forces are strongest -- may be less massive, which would provide indirect evidence for tidal stripping.

Given the insufficient signal-to-noise (S/N) of weak lensing, correlations between subhalos within the same host provide a promising alternative way to measure tidal stripping, as demonstrated by \citet{chamberlain15} using subhalos in N-body simulations. \citet{cohn12} and \citet{cohn14} had shown the existence of such correlations and discussed a variety of consequences in physical and velocity space. 
\citet{chamberlain15} pointed out that subhalo-subhalo correlations can be used to address an important open question in galaxy formation: did the star formation of satellites galaxies end upon accretion?
The striking color difference between galaxies in cluster and field environments -- satellites are mostly ``red and dead'' while field galaxies are mostly blue and actively forming stars -- has prompted a search for the mechanisms within clusters that end star formation in satellites.
These include accretion shocks \citep{balogh00, dekel06}, strangulation \citep{larson80},  ram-pressure stripping \citep{gunn72, abadi99}, and the effect of many high-speed encounters with other satellites  \citep{farouki81, moore96} -- all of which act to either forcibly remove gas from the subhalo or prevent it from cooling enough to form stars.

It is possible to reproduce many observed statistics of quiescent galaxies using models relying upon quenching of star formation caused by intra-cluster processes.  For example, \citet{wetzel13} used a galaxy group catalog from SDSS Data Release 7, combined with the quiescent fractions from COSMOS survey, and a cosmological N-body simulation to study the star formation histories of satellite galaxies at $z \approx 0$.
They found that satellite quenching is consistent with the statistics of quiescent galaxies if quenching is a `delayed-then-rapid' process: the satellites remain actively star forming for $2-4$ Gyr after their first infall, unaffected by the host halo, after which the quenching occurs rapidly with an SFR e-folding time $< 0.8$ Gyr.

Alternatively, it is also possible that other processes besides satellite quenching may be dominant in determining quiescent fractions in clusters.  For example,  the age-matching model of \citet{hearin13} can reproduce many measurements of galaxy luminosity, color, clustering, and weak lensing \citep{hearin14, watson15} without relying on satellite quenching processes.
By placing the reddest galaxies in the oldest dark matter halos -- which tend to be the ones that formed in denser environments -- the \citet{hearin13} model naturally reproduces the large red fraction seen in observations.

However, recently \citet{zu15a} showed that the original age-matching model of \citet{hearin13} is in tension with measurements of the halo mass of isolated blue galaxies \citep{mandelbaum16}.
\citet{zu15a} study two quenching models and show that a model in which halo mass alone determines quenching fits the SDSS measurements \citep{zu15b, mandelbaum16} better than a hybrid model which depends on stellar mass and host halo mass (for satellites).
While both models do well at fitting red galaxy clustering and lensing, the halo quenching model is much better at modeling the halo mass for massive blue central galaxies.  
\citet{zu15a}  show that the color dependence of the $\langle M_h | M_*\rangle$ relation in the publicly available mock galaxy catalogs of \citet{hearin14} is inconsistent with SDSS measurements, since this mock tends to place blue and red centrals of similar $M_*$ in halos of similar mass, which is heavily disfavored by the data.  It is unclear whether this conclusion can be generalized to the broader class of age-matching models, or if it is specific to the original construction of \citet{hearin13}, since \citet{campbell16} argue that a slightly modified parametrization of age-matching fits the measurements of \citet{zu15a}.

The detection of the splashback feature by \citet{more16} provides additional insight into quenching processes in clusters.  \citeauthor{more16} detect the splashback feature in both red and blue galaxies.  This shows that galaxies can complete at least one full orbit within their host clusters while remaining unquenched (i.e.\ blue).  Given that orbital times in the outskirts of halos can be a large fraction of the Hubble time (e.g., 6 Gyr), this result may pose a challenge for the model of \citet{wetzel13}.  \citeauthor{more16} also find that the red fraction of cluster galaxies exhibits a pronounced feature at the splashback radius $r_{\rm sp}$.  While the naive interpretation of this sharp feature at the splashback radius is that crossing $r_{\rm sp}$ modifies galaxy colors (i.e.\ host-quenching), similar behavior also naturally arises in age-matching models without host-quenching.  In those models, the sharp feature in the red fraction is a consequence of the sharp transition at the splashback radius from the 2-halo region to the 1-halo region.  

In light of these conflicting models, we carry out the subhalo-subhalo clustering measurements proposed in \citet{chamberlain15} in order to provide model-independent constraints on the relationship between subhalo infall and quenching times.
Alternative models of galaxy formation may then be distinguished based on their predictions for the fraction of quiescient satellites that quenched upon accretion, rather than while isolated. We assume a flat universe and $\Omega_m$ = 0.3.
In Section 2 we describe our SDSS data samples.
In Section 3 we describe our method of the measurement.
In Section 4 we describe measurement results of subhalo correlations and compare infall, dynamical, and quenching timescales for red SDSS satellites.
In Section 5 we discuss possible systematics as well as implications of our measurements for models of galaxy formation.

\section{Data}
\label{sec:data}

For our subhalo sample we use members of SDSS redMaPPer clusters \citep{rykoff14, rozo15a}.
We use clusters with redshift $0.15 < z < 0.41$ and more than ten member galaxies, i.e., we require the richness $\lambda > 10$ (note that the public catalog only goes down to $\lambda=20$).
We also require the clusters to have a high redMaPPer centering probability, \texttt{Pcen} > 0.8.
These cuts reduce the number of clusters to 11,800.
In addition to the cuts on cluster properties, we only use satellites with a membership probability \texttt{Pmem} > 0.8, ensuring that our satellite sample is pure.
\citet{rozo15a} has shown that the photometric cluster selection done by redMaPPer is effectively as good as a spectroscopic selection for those satellites with a high membership probability.  We will be interested in studying the satellite clustering as a function of distance from the center of the cluster, $r_c$, so for this purpose we define three bins $r_c = 0.1-0.3\mpch$, $r_c = 0.3-0.6\mpch$, and $r_c = 0.6-0.9\mpch$.
The range in $r_c$ for each bin was chosen to obtain sufficient signal-to-noise (S/N) in all three bins: each bin of increasing $r_c$ has 110144, 42742, and 6728 galaxies.

We cross-correlate these subhalos with the redMaGiC galaxy catalog \citep{rozo15b}.
We use \redmagic{} because many of these bright red galaxies are members of the redMaPPer clusters and may compose infalling groups with the redMaPPer members we study.
We do not simply use the redMaPPer satellites because redMaPPer imposes a spatial selection around each cluster, rejecting all galaxies outside a radius $R_\lambda = 1 \, {\rm Mpc}/h \, (\lambda/100)^{0.2}$.
This might introduce edge effects in our measurement so we use the spatially uniform \redmagic{} sample for cross-correlations (see Sec.~\ref{sec:results} for details).
The \redmagic{} algorithm selects only red galaxies with good photometric redshifts with a median bias $z_{\rm spec} - z_{\rm photo}$ of 0.005 and scatter $\sigma_z / (1+z)$ of 0.017.
When performing cross-correlations we select only redMaGiC galaxies within 0.02 of the cluster redshift.

For each \redmapper{} member, we use SED fits to the broadband photometry to determine the age, or time since the onset of star formation.
The prior on star formation history is taken to be delayed tau models \citep{maraston10}, which produce a continuous but exponentially declining star formation history.
We are most interested in when the subhalo quenched, or stopped forming stars.
Due to the exponential decline, it is not a bad approximation to take this measure of age as the quenching time,
$\tquench \equiv (\int \de t \; t \; {\rm SFR}(t)) / \int \de t \; {\rm SFR}(t)$,
where we have used the star formation rate (SFR)-weighted age.
The initial mass function is taken from \citet{salpeter55} over the range $0.1 - 100 \, M_\odot$.
See Appendix A of \citet{moustakas13} for more details on the SED modeling.
In Sec.~\ref{sec:galform} we discuss how this definition of $\tquench$ connects with definitions used in galaxy formation models.

\begin{figure}
\resizebox{85mm}{!}{\includegraphics{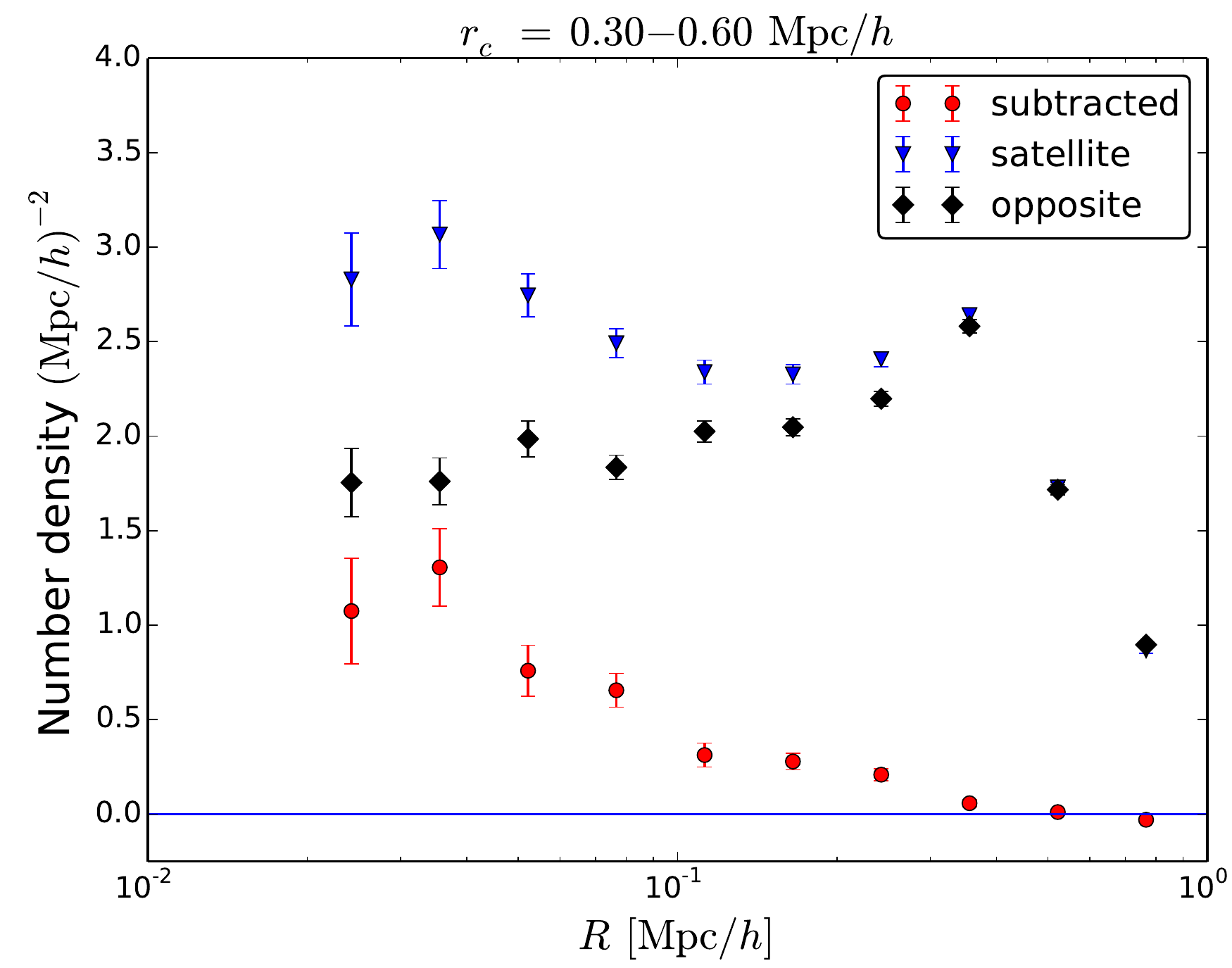}}
\caption{Number density of redMaGiC galaxies around redMaPPer members (blue points) and around their opposite positions relative to the cluster center (black points).
The difference (red points) isolates the subhalo clustering signal.}
\label{fig:method}
\end{figure}

\section{Method}
\label{sec:method}

\citet{chamberlain15} used simulations to split subhalos between those that were recently accreted and those accreted more than one dynamical time ago.
The dynamical time was estimated as the period for a circular orbit at radius equal to the distance from satellite to its central galaxy.
The results are shown in their Figure 2: the oldest subset showed no significant correlation at large radii. In contrast, the subhalos which infell recently did show large-scale correlations.
This means the measurement of large-scale correlations can be used to determine when subhalos fell into their hosts.

We seek to detect these large-scale correlations in data by cross-correlating subhalos (redMaPPer members) with the \redmagic{} sample by counting the number of \redmagic{} galaxies in annuli centered on each \redmapper{} subhalo.
However, since subhalos are by definition always part of a larger halo, the raw number counts measured around subhalos will be dominated by the host.
In order to remove this host signal, we follow the method of \citet{pastor11} and \citet{chamberlain15}, subtracting the counts around the subhalo's ``mirror'' point reflected about the cluster center.
Since this point is at the same distance $r_c$ from the cluster center, if the host cluster profile is spherically symmetric it will have the same contribution from the host.
This method is illustrated in Fig.~\ref{fig:method}, which shows the counts around the redMaPPer satellite and the opposite point.
Both are dominated by the mis-centered host signal, but after subtraction the subhalo clustering signal is clearly visible at small scales and falls smoothly to zero at large scales.

This subtracted signal is our estimator used in all remaining plots: we count \redmagic{} galaxies in annuli centered on each \redmapper{} subhalo, subtract the counts of \redmagic{} galaxies around each mirror point, and divide by the number of subhalos.
For all measurements, we divide the SDSS data into 200 spatially uniform patches and estimate a jackknife covariance by removing each patch in turn \citep{norberg09}.
All figures show the diagonal $1\sigma$ jackknife errors while our signal-to-noise (S/N) estimates incorporate the full jackknife covariance.

Note that we measure correlations in 2D, projected bins of cluster-centric distance $r_c$.
This is in contrast to Figure 2 of \citet{chamberlain15}, which shows 3D correlations.
In Sec.~\ref{sec:syst} we argue that the qualitative conclusions we draw from the 2D correlations should be similar to those based on 3D correlations. Moreover in future work we intend to present correlations in deprojected 3D cluster-centric distance bins to compare more directly with \citet{chamberlain15}.

\begin{figure}
\resizebox{85mm}{!}{\includegraphics{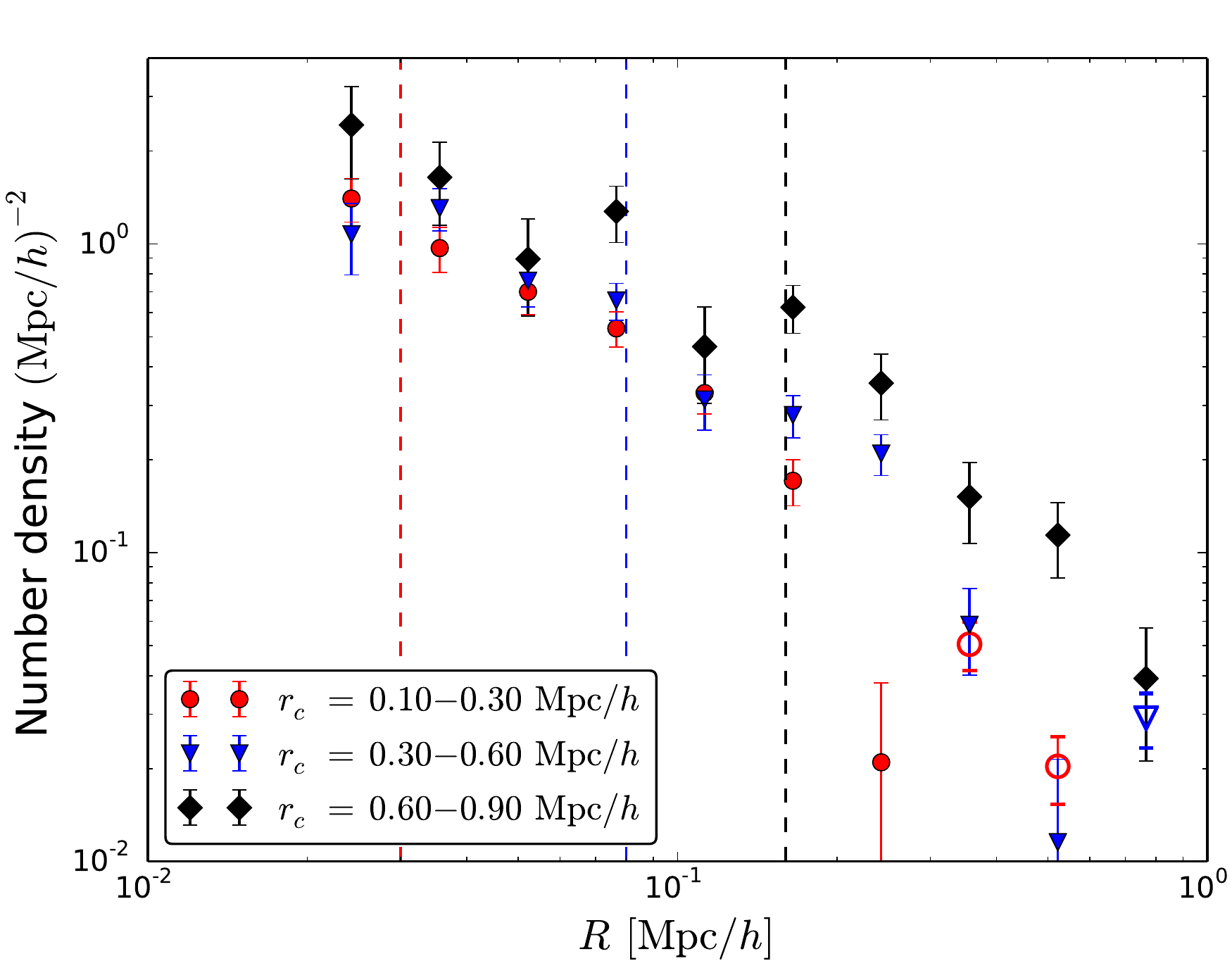}}
\caption{Projected number density of redMaGiC galaxies stacked around subhalos, for all 3 bins of cluster-centric distance $r_c$.
Above $\sim 0.1 \mpch$ there is a clear trend of increasing correlations for larger cluster-centric distance.
Vertical lines show the estimated tidal radius for each subhalo sample: all measurements show signficant correlations with unbound material beyond the tidal radius.}
\label{fig:full_sample}
\end{figure}

\begin{figure}
\resizebox{85mm}{!}{\includegraphics{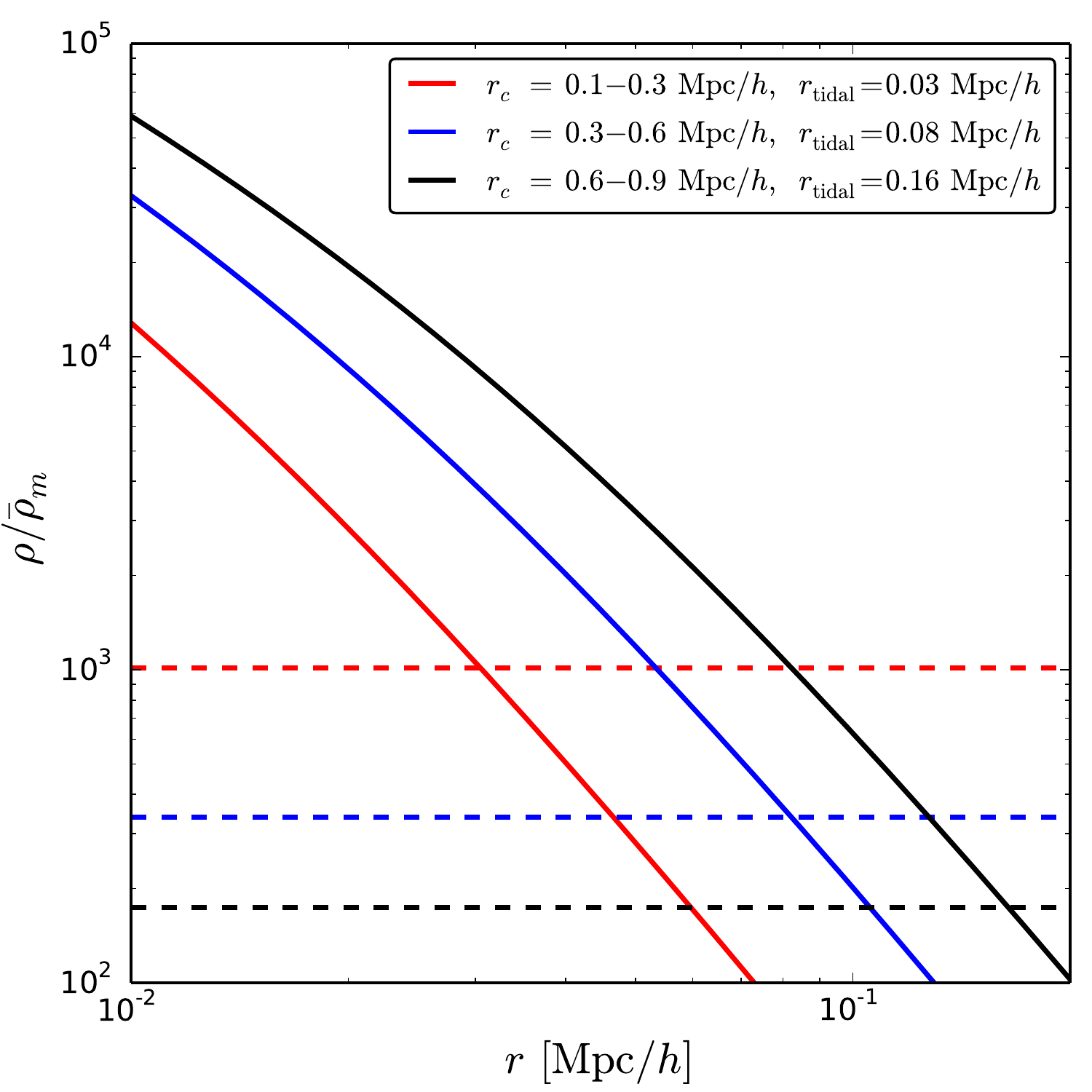}}
\caption{The subhalo profile (solid lines) and background host density (dashed lines) for each $r_c$ bin.
The estimated $\rtidal$ is the radius where the two density profiles cross.
}
\label{fig:rtidal}
\end{figure}

\section{Results}
\label{sec:results}

\subsection{Correlations beyond the tidal radius}
\label{sec:full}

In Fig.~\ref{fig:full_sample} we show the results for three cluster-centric radius bins, $0.1 < r_c < 0.3$, $0.3 < r_c < 0.6$, and $0.6 < r_c < 0.9 \mpch$.
The correlations are clearly detected in each case, and fall as $\sim 1/R$ with increasing scale.
Note there is an increase in correlations for larger $r_c$ bins which may be due partly to the positive correlation of $r_c$ with stellar mass (see Fig.~\ref{fig:mstar}).
In order to interpret these measurements, we next estimate the subhalos' tidal radius, beyond which unbound material can be stripped by tidal forces from the host.

Measuring correlations well beyond the tidal radius would indicate that at least some of the subhalos were recently accreted, since otherwise unbound material would have been stripped.
The tidal radius or Hill radius is roughly the radius where $\rhosub \sim \rhohost$, where the dynamical time for particles orbiting the subhalo becomes comparable to the dynamical time for the subhalo to orbit its host.
To obtain $\rhosub$ we use the results of \citet{li15}, which measured lensing masses for a subset of our subhalos, those which overlap with the CFHT Stripe-82 Survey.
\citet{li15} found best-fit subhalo masses $10^{11.3}$, $10^{12.0}$, and $10^{12.5} M_\odot/h$ for increasing distance from the cluster, using the same three $r_c$ bins as our measurements.
In Fig.~\ref{fig:rtidal} we show NFW density profiles \citep{nfw1997} for the three bins using these masses.

To obtain the average local host density at the satellite locations, we use the mass-richness relation of \citet{simet16}:
\begin{equation}
M_{\rm 200} = 10^{14.344} M_\odot/h \times {(\lambda / 40)^{1.33}} \, .
\end{equation}
We then estimate $\rhohost(r)$ using this host mass and an NFW profile, and plot the result in Fig.~\ref{fig:rtidal}. Here $r$ is the 3D cluster-centric distance. By assuming the host halo is spherically symmetric, and the distribution of satellites follows an NFW form, $r$ can be estimated by
\begin{equation} \label{eq:3drc}
r(r_c) = \frac{\int_{0}^{\sqrt{{R_{200}}^2-r_c^2}}\rho_{\rm nfw}(\sqrt{r_c^2+x^2}) \times \sqrt{x^2 + r_c^2} \, {\rm d}x}{\int_{0}^{\sqrt{{R_{200}}^2-r_c^2}}\rho_{\rm nfw}(\sqrt{r_c^2+x^2}) \,{\rm d} x} \, .
\end{equation}
The host and subhalo density profiles cross at $\rtidal \sim 0.03$, $0.07$, and $0.15\mpch$ for the three bins of increasing $r_c$.
A rough comparison to the simulation results of \citet{chamberlain15} shows that our results are comparable: for their $0.4 < r_c < 0.6 \mpch$ bin, the local host density and subhalo density match at $\sim 0.045 \mpch$ (see Fig. 1 of that work).
This is slightly smaller but close to the tidal radius $\sim 0.07 \mpch$ that we find for subhalos between $0.3 < r_c < 0.6 \mpch$; exact agreement is not expected since we have not attempted a comparison with matched host halo masses. 

Material at larger radii should be unbound, so we look for correlations beyond $\rtidal \sim 0.03$, $0.07$, and $0.15\mpch$, for each of the three $r_c$ bins.
Comparing these scales to Fig.~\ref{fig:full_sample}, we do see correlations well beyond $\rtidal$ in each case. We find the S/N using points in the range $\rtidal$ to $r_c/2$ is 11.1, 10.3, and 6.1 corresponding to the three bins of increasing $r_c$.
Thus all show significant detections.
This implies that some of the subhalos recently fell into their hosts and there has not been enough time for unbound material to be torn away, i.e., $\tinfall < \tdyn$, where $\tinfall$ is the infall time and $\tdyn$ is the dynamical time.
In Table~\ref{table:sn_full} we also show the S/N using a wider range of scales $\rtidal < R < r_c$, but note that
points close to $r_c$ may be subject to several systematics discussed in Section~\ref{sec:syst}.
In the following section we estimate the dynamical time and compare it to the quenching timescale for the subhalos.

\begin{table}
\centerline{
\small
\begin{tabular}{| l | c | c |}
\hline
$r_c$ (Mpc/$h$) & \multicolumn{2}{c}{S/N} \\ \hline
                         &[$\rtidal$, $r_c$/2] & [$\rtidal$, $r_c$] \\ \hline
0.1 - 0.3 & 11.1 & 14.7 \\ \hline
0.3 - 0.6 & 7.1  & 8.8 \\ \hline
0.6 - 0.9 & 6.1  &  6.8  \\ \hline
\end{tabular}
}
\caption{Signal-to-noise of the correlations beyond the tidal radius shown in Fig.~\ref{fig:full_sample}.
The measurement is significant regardless of cluster-centric distance $r_c$.}
\label{table:sn_full}
\end{table}

\subsection{Dynamical and quenching timescales}

\begin{figure}
\resizebox{85mm}{!}{\includegraphics{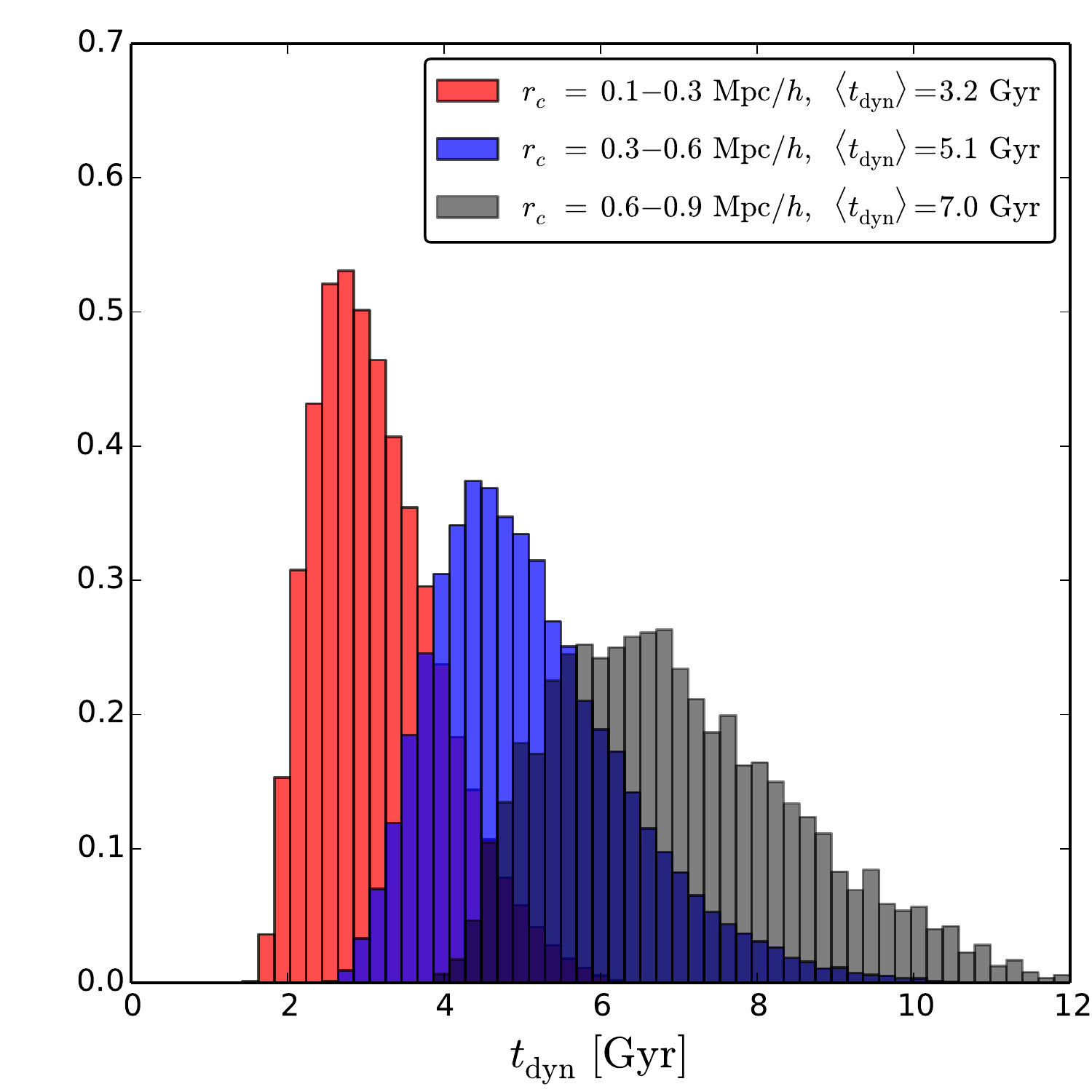}}
\caption{Distribution of dynamical times of the subhalos for each $r_c$ bin. The dynamical time is estimated using Eq.~(\ref{eq:tdyn}). Here we use the redMaPPer mass richness relation and assume an NFW profile to estimate the host mass. It can be seen that the dynamical time is increasing with $r_c$ with the mean value for the outer most bin $t_{dyn}$ $\sim 6$ Gyr.}
\label{fig:tdyn}
\end{figure}

\begin{figure}
\resizebox{85mm}{!}{\includegraphics{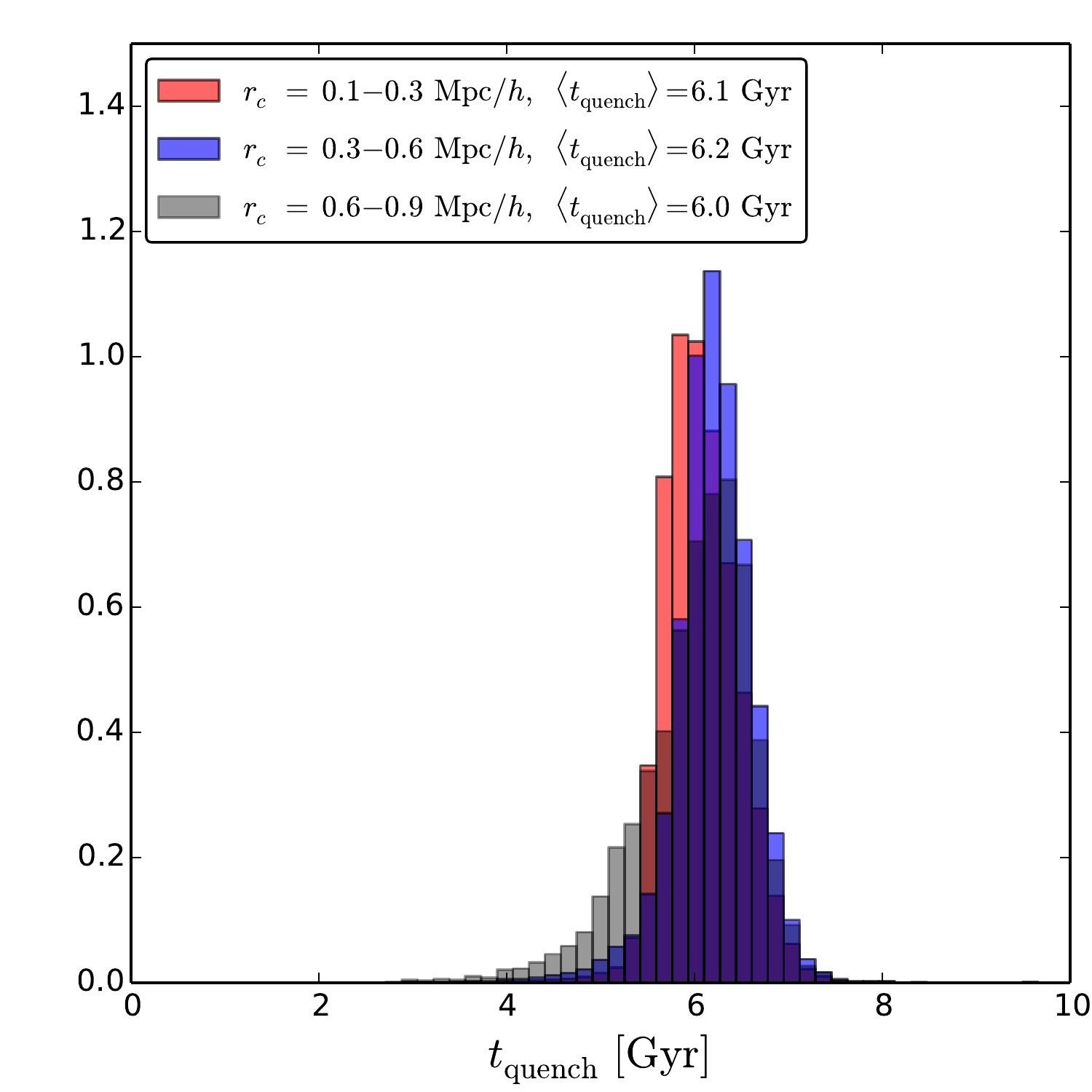}}
\caption{Quenching time since the redMaPPer subhalos stopped forming stars. For all three $r_c$ bins, the mean quenching time is $\sim 6$ Gyr. Thus for the inner two $r_c$ bins we have $\tdyn < \tquench$.
}
\label{fig:sfrage}
\end{figure}

We can estimate the dynamical time for each subhalo using its distance from the cluster center $r_c$ and the host density profile.
As in Sec.~\ref{sec:full}, we estimate the host density using the redMaPPer mass richness relation and assume an NFW profile.
The dynamical time is then
\begin{equation} \label{eq:tdyn}
\tdyn = 2\pi \sqrt{r^3 / (G M(< r))} \, ,
\end{equation}
where $M(< r)$ is the host mass contained within the 3D radius $r$, which is estimated by Eq.~(\ref{eq:3drc}).
In Fig.~\ref{fig:tdyn} we show histograms of the resulting dynamical times, for the three $r_c$ bins.
The mean is small for the innermost bin, $\sim 3$ Gyr, increasing to $\sim 6$ Gyr for the galaxies farthest from the cluster center.
This is expected due to the smaller density at larger distances from the cluster center.

We next compare these dynamical times to the quenching time determined from SED fits (see details in Sec.~\ref{sec:data}).
In Fig.~\ref{fig:sfrage}, we show the distribution of $\tquench$ using the mean of the posterior for each subhalo.
The distribution peaks at $\sim 6$ Gyr, a value greater than the dynamical times $\sim$ 3-4 Gyr (see Fig.~\ref{fig:tdyn}) for subhalos with $r_c < 0.6 \mpch$.
Thus many of our subhalos have $\tinfall < \tdyn < \tquench$, implying that these galaxies ceased star formation long before being accreted by their hosts.

This would be inconsistent with models of star formation in which these subhalos were actively forming stars up until the time they fell into their current hosts.
In such models, infall into the cluster leads to stripping of the gas and quenching of star formation.
Hence such models require that $\tquench < \tinfall$ for a large fraction of our subhalos.

\subsection{Other tests}

\begin{figure*}
\centering
\resizebox{58mm}{!}{\includegraphics{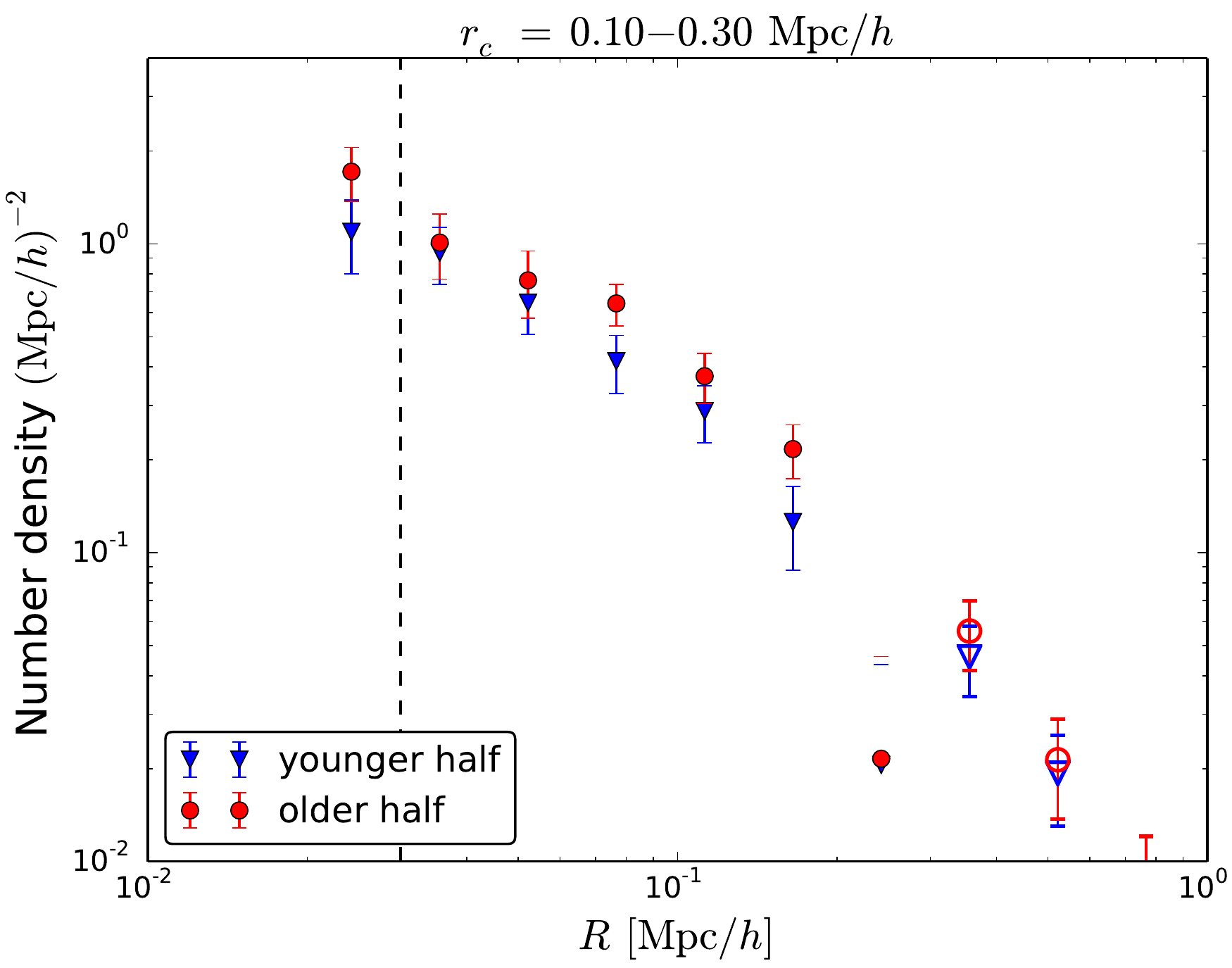}}
\resizebox{58mm}{!}{\includegraphics{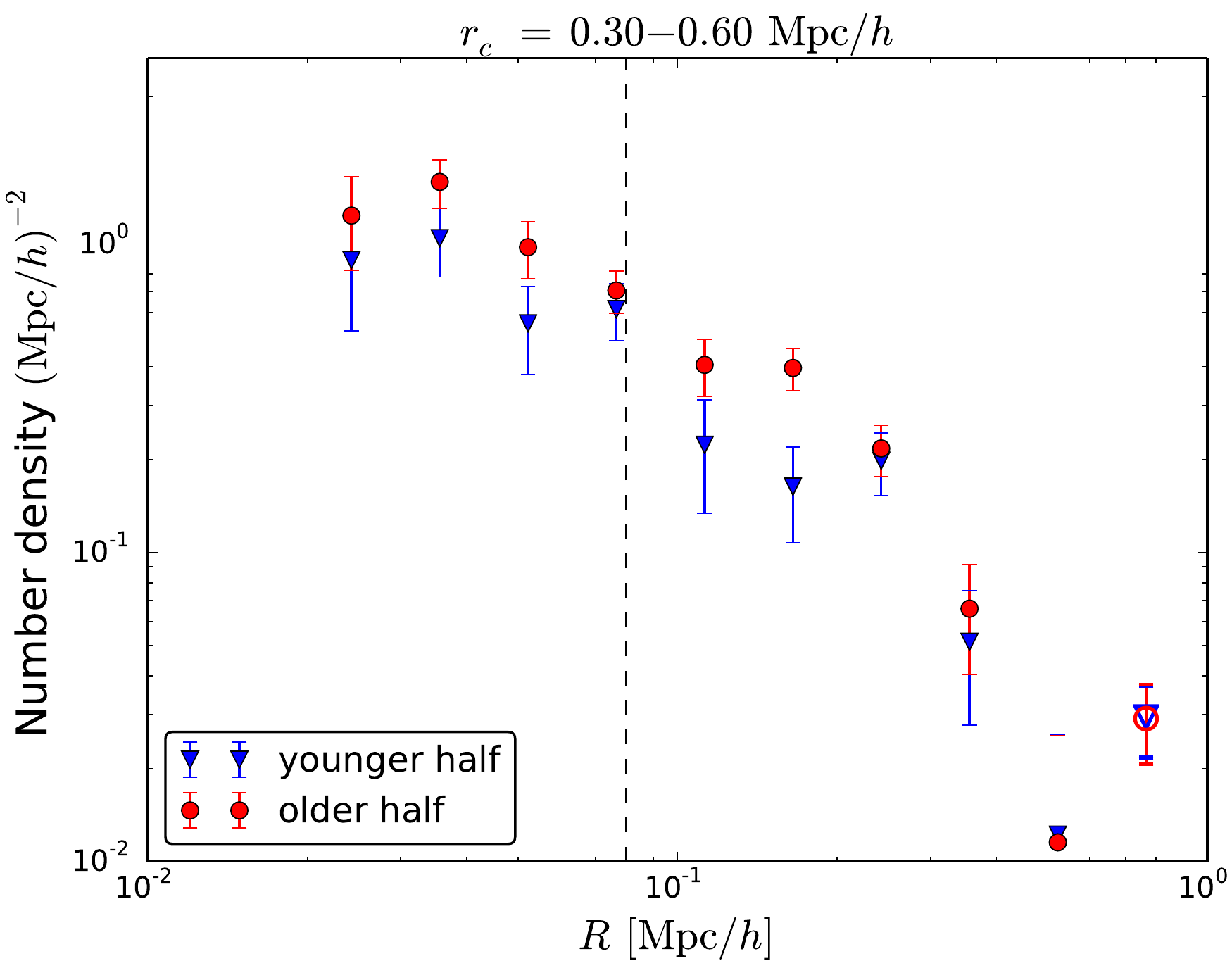}}
\resizebox{58mm}{!}{\includegraphics{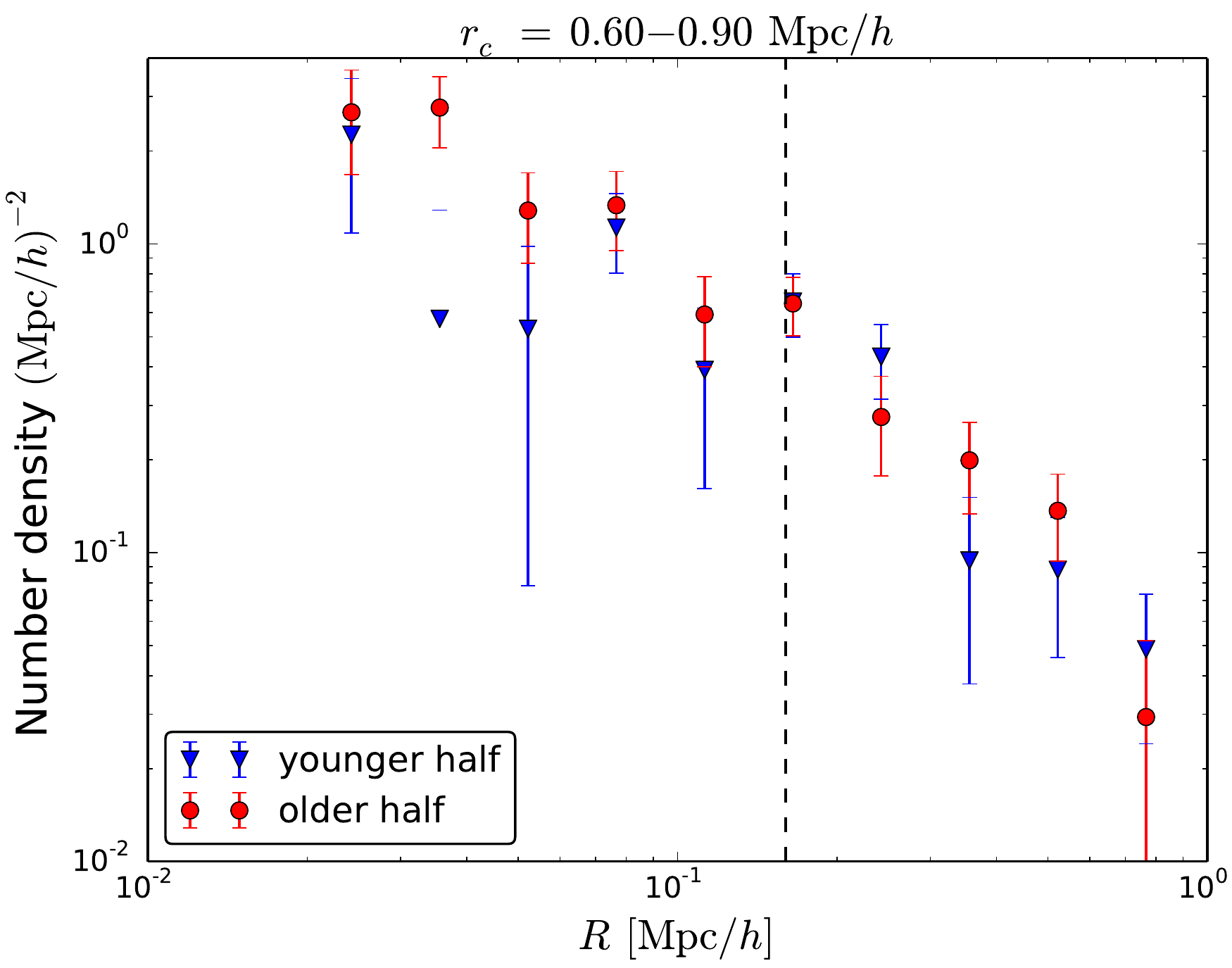}}
\caption{Comparison between correlations of older (red circles) and younger (blue triangles) subhalos, as determined from SED fit results for $\tquench$.
Each panel shows results for subhalos at different distances $r_c$ from the host cluster.
Subhalos with $r_c < 0.6 \mpch$ show correlations well beyond the tidal radius (vertical dashed line).
These long-range correlations are present for both the young and old samples.}
\label{fig:age_split}
\end{figure*}

\begin{figure*}
\centering
\resizebox{58mm}{!}{\includegraphics{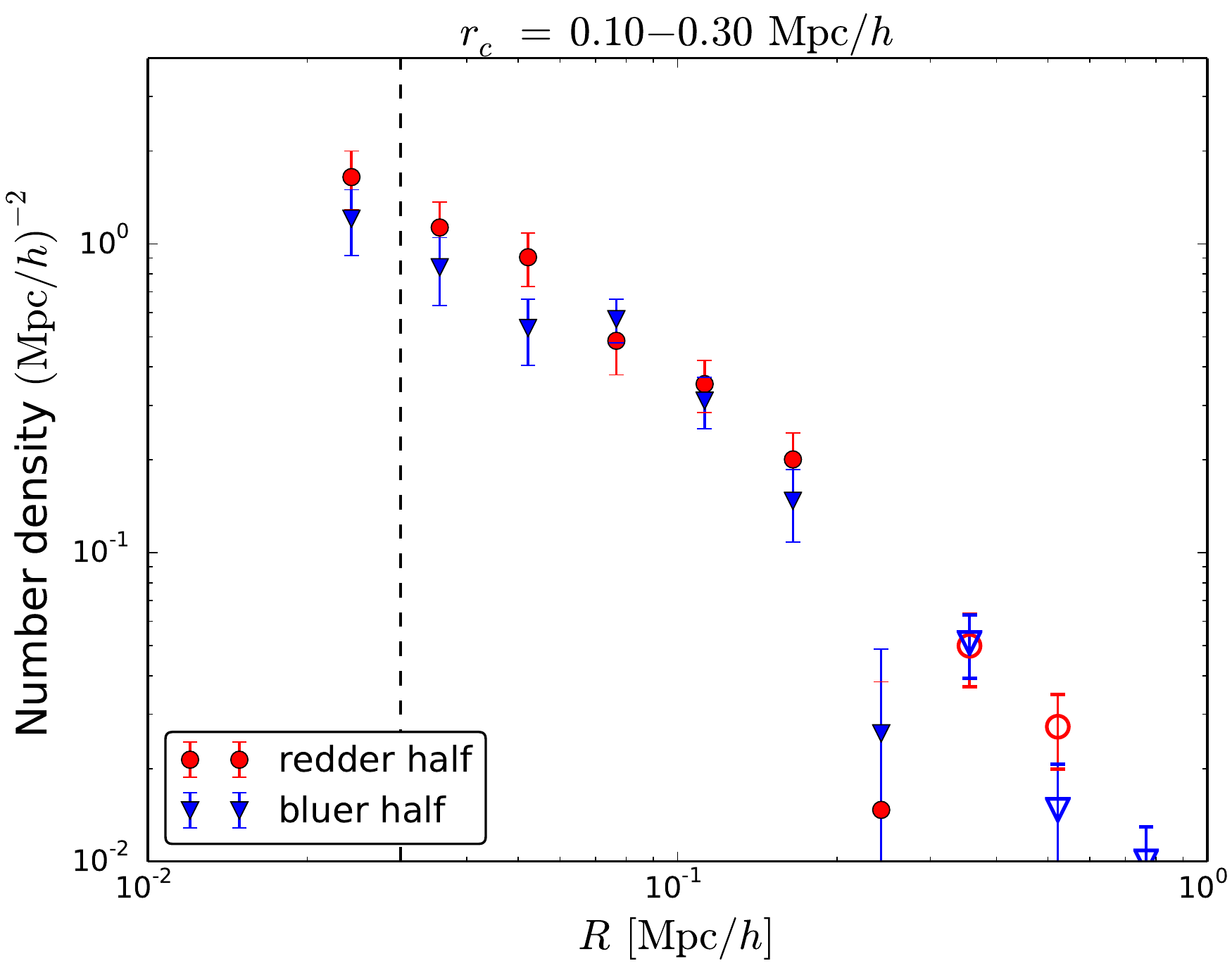}}
\resizebox{58mm}{!}{\includegraphics{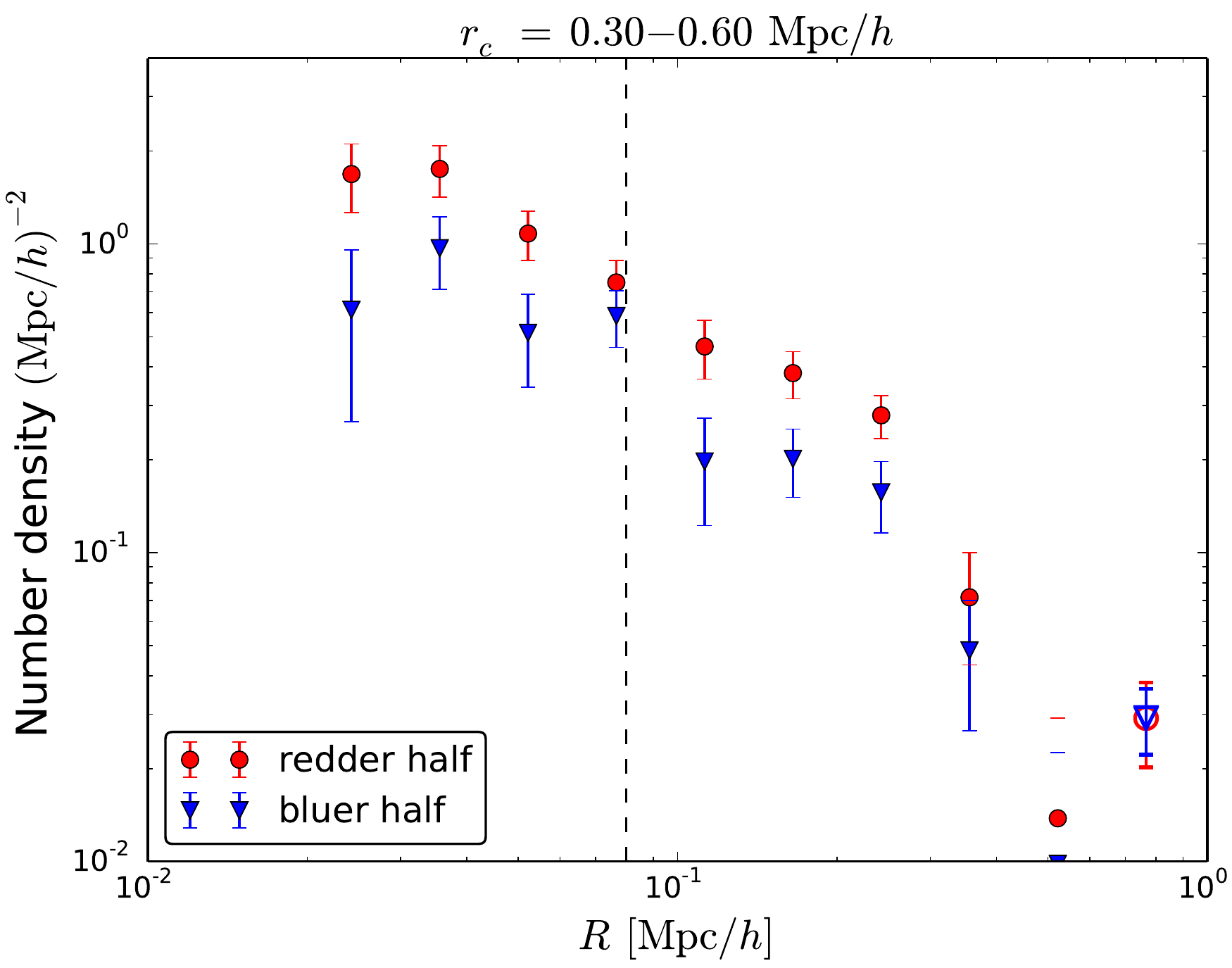}}
\resizebox{58mm}{!}{\includegraphics{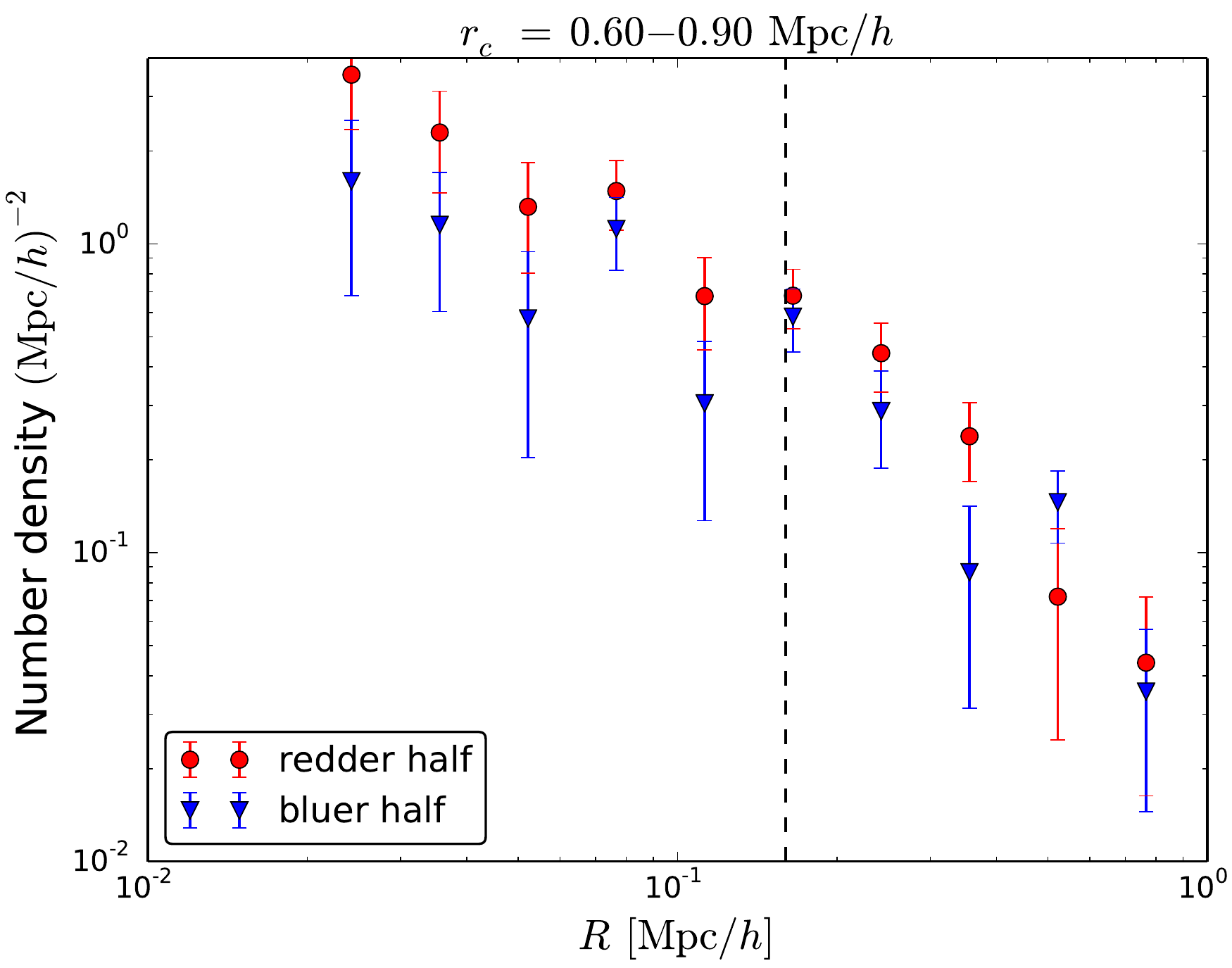}}
\caption{Same as Fig.~\ref{fig:age_split} but showing subhalos split by color: those redder (red circles) and bluer (blue triangle) than the red sequence model.
Since redder galaxies have been quiescent for longer, these results are similar to the $\tquench$ split but less dependent on detailed SED modeling.
}
\label{fig:color_split}
\end{figure*}

As an additional test of the quenching scenario, we have redone the measurement using the older half of subhalos in Fig.~\ref{fig:sfrage}, most of which have $\tquench > 6$ Gyr.
The results are shown in Fig.~\ref{fig:age_split}, and again we see strong correlations well beyond the tidal radius. 
For a more model-independent test of age we split the subhalos into halves based on color, a proxy for age since the reddest, deadest subhalos are the ones that ceased forming stars long ago.
In Fig.~\ref{fig:color_split} we show the correlations measured around the ``redder'' subhalos, defined as those which are redder than the redMaPPer red sequence model at that redshift.
Again we see strong correlations beyond the tidal radius, for all $r_c$ bins. 
Thus, many of these subhalos recently fell into their hosts, long after they had ceased forming stars.

We also tested the dependence of our results on the specific galaxy samples by measuring the auto-correlation of redMaPPer satellites.
This gave similar results with correlations persisting well beyond the tidal radius.
The redMaPPer satellites are dimmer than redMaGiC with a cut at $L > 0.2 L_\ast$ instead of $L > 0.5 L_\ast$. But redMaPPer also enforces a spatial selection around the cluster center, which may bias the redMaPPer auto-correlation for the larger $r_c$ bins.
For this reason we did not choose the auto-correlation for our fiducial measurements.

\section{Discussion}

We have shown that \redmagic\ galaxies are significantly correlated with \redmapper\ satellites, on angular scales exceeding the estimated tidal radii for satellites at small cluster-centric distances.  This result holds even when we subselect the reddest satellites, or those believed to have been quenched for the longest times.  Our measurements can be compared with the simulation results of Chamberlain et al (2015), and suggest that a significant fraction of the massive satellites in our sample were quenched prior to infall into their host cluster halos. Future studies with ongoing surveys using the clustering of satellite galaxies and their weak lensing signal can shed light on galaxy formation models and  probe possible interactions of dark matter.  Below, we discuss potential systematics in our measurements, and then discuss implications for satellite quenching models.

\subsection{Possible systematics}
\label{sec:syst}

First, we discuss possible systematics in our measurement of correlations beyond the tidal radius.
Two effects which we have neglected will become increasingly important at projected distances $\sim r_c$, i.e., near and beyond the cluster center.
The first is over-subtraction of the subhalo profile.
When we measure correlations around the mirror point (see Sec.~\ref{sec:method} and Fig.~\ref{fig:method}) there will be some positive contribution from the subhalo, the magnitude of which will become important at scales $R \sim 2\, r_c$ when the annulus $R$ contains the subhalo.  Note that this systematic contributes with negative amplitude due to the subtraction and so does not weaken our case for the detection of subhalo correlations beyond the tidal radius.

A second effect could result from systematic cluster miscentering.
If the true cluster center is on average closer to the subhalo than redMaPPer's most probable center, then the subtraction of the mirror point signal will imperfectly remove correlations due to the host cluster.
This  would contribute most strongly at larger scales $R \sim r_c$: as seen in Fig~\ref{fig:method}, the cluster signal as measured around the opposite point is roughly flat within $R \sim r_c / 2$, then begins rising to peak at $R \sim r_c$ before decreasing again.
Thus residuals of the subtracted cluster signal could cause a roughly constant systematic below $r_c / 2$ and increase in magnitude at larger scales. We found no evidence of such a trend in our measurements. 
Note that we only use clusters with a center that is both (i) the most likely center for that cluster and (ii) truly looks like a central: these are galaxies which are bright, red, and near the center of the spatial distribution of cluster members. 
See \citet{rykoff14} for more information on the redMaGiC centering filter and \citet{rozo14} for further tests of redMaPPer centering, including comparison to X-ray centers for a subset of clusters.

Due to imprecise line-of-sight information on galaxy positions, some fraction of the \redmapper{} members may lie outside the halo.
Assuming this interloper component has a constant spatial density, it will be most important at large $r_c$ since the profile of actual cluster galaxies drops with distance from the cluster center.
According to \citep{rozo15a}, the fraction of such interlopers is estimated to be 6\% averaged over the whole cluster.
If interlopers were responsible for much of the signal we detect, the measured angular correlation  for neighbors around \redmapper\ galaxies at small $r_c$ would be $\sim 16\times$ smaller than the analogous correlation function at $r_c \gg r_{\rm vir}$. We do not find this, so conclude that interlopers are not a significant factor in our analysis. 

The previous paragraph discussed the effect of projections of distant interlopers, next we discuss the effect of two types of less catastrophic projections.
First, we measure correlations with redMaGiC galaxies in 2D projected distance (i.e., the x-axis of Fig.~\ref{fig:full_sample} is a projected quantity).
\citet{chamberlain15} showed that results for subhalo-subhalo correlations in 3D and in projection are qualitatively similar. For example, in their Figures 3 and 4, they show that 3D and projected correlations show the same features between different cluster-centric distance bins.
Thus, we do not worry further about this type of projection effect.
Second, we use 2D cluster-centric bins $r_c$, meaning that each $r_c$ bin contains subhalos over a range of larger 3D distances.
In this paper, we have partially accounted for this effect by estimating the average 3D cluster-centric distance via Eq.~(\ref{eq:3drc}), and using that mean 3D quantity to estimate $\rtidal$ and $\tdyn$.
In addition, we have carried out a deprojection of our angular correlations for the two lower $r_c$ bins; preliminary results suggest that the long-range correlations persist.
We will present a quantitative comparison of our deprojected correlations with galaxy formation models in a separate study.

\subsection{Implications for models of galaxy formation}
\label{sec:galform}

\begin{figure}
\resizebox{85mm}{!}{\includegraphics{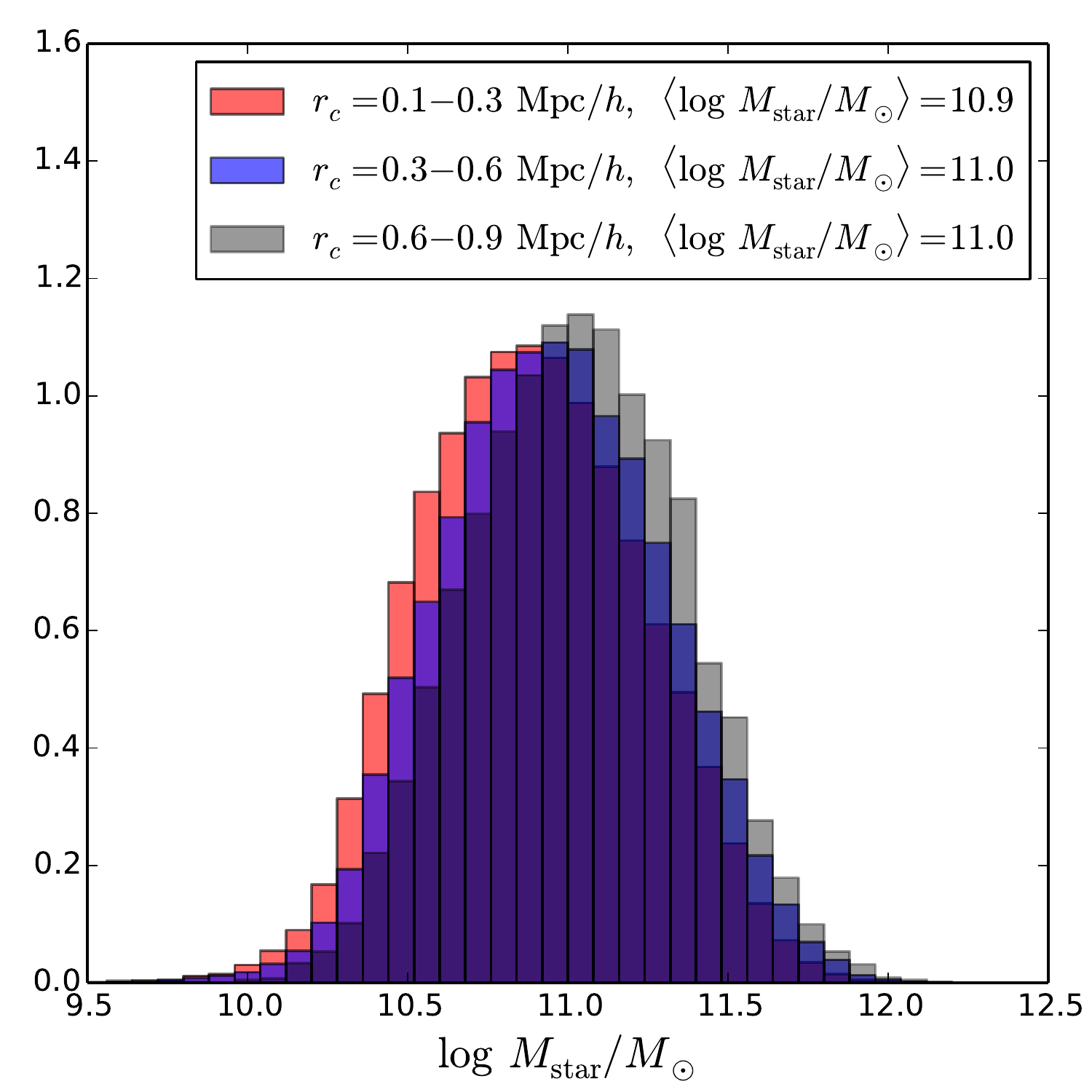}}
\caption{Normalized stellar mass distributions for the three cluster distance bins.
}
\label{fig:mstar}
\end{figure}

\begin{figure*}
\centering
\resizebox{78mm}{!}{\includegraphics{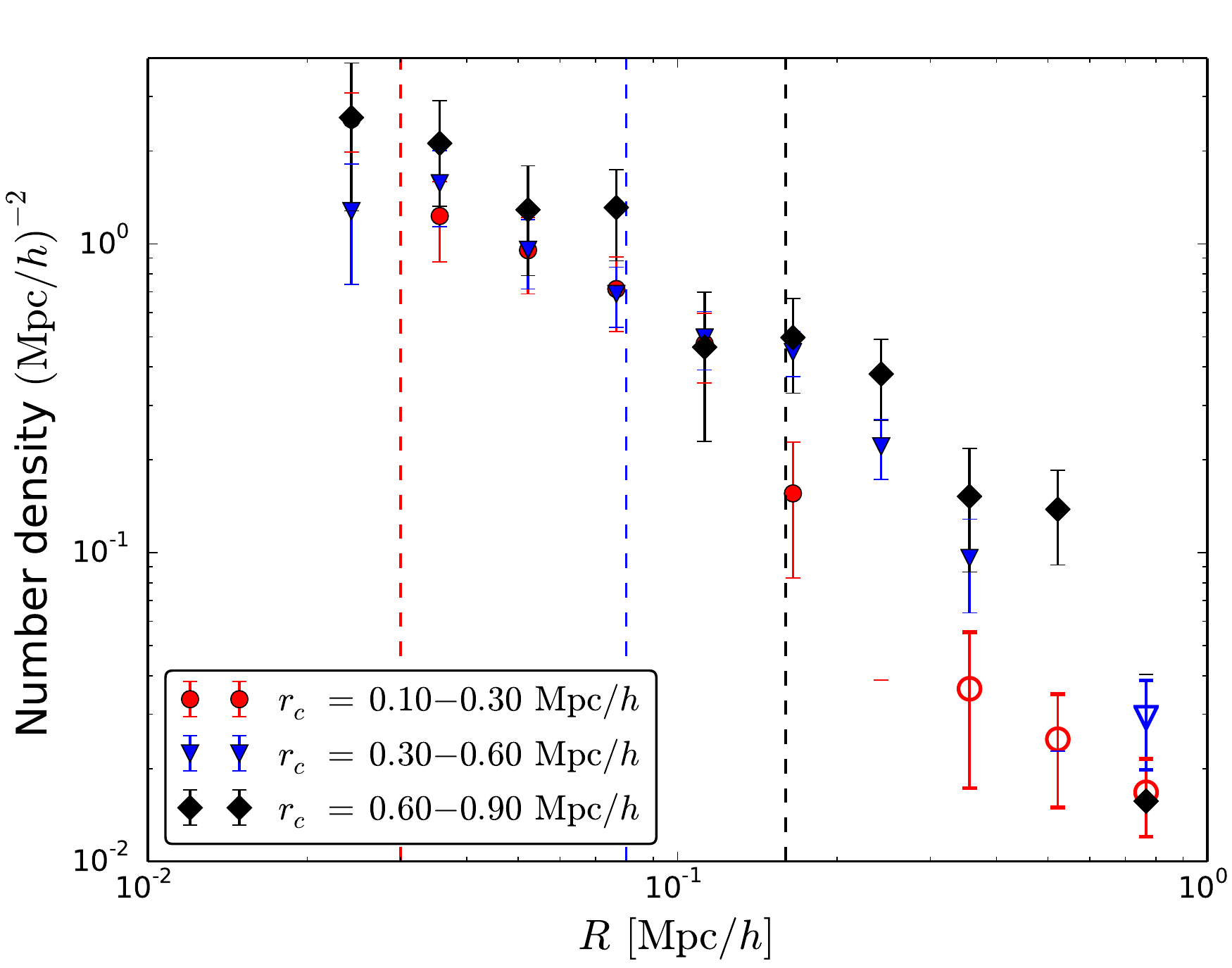}}
\resizebox{78mm}{!}{\includegraphics{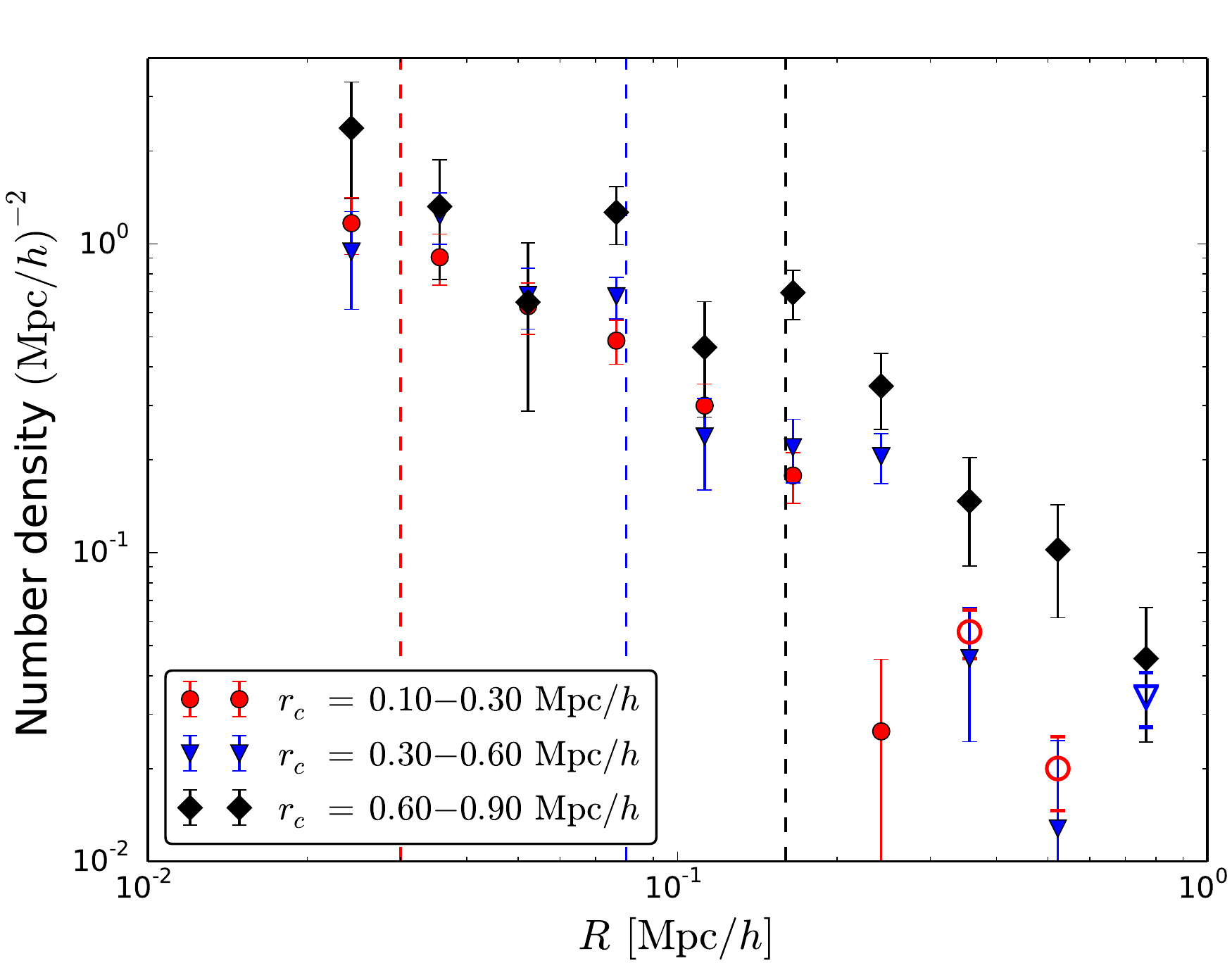}}
\caption{Projected number density as in Fig.~2, but split by satellite stellar mass:  higher (left panel) and lower (right panel) than $10^{11}M_\odot /h$ .
 }
\label{fig:stellarmass_split}
\end{figure*}

We have argued that the absence of tidal stripping favors isolated quenching of a significant fraction massive galaxies, prior to their infall into galaxy clusters. 
Here we compare our results to host-quenched fractions predicted by more specific models of galaxy formation.
The model of \citet{wetzel13} predicts a strong dependence of the host-quenched fraction with subhalo stellar mass and host halo mass.
Our clusters have mass $\gtrsim 10^{14} M_\odot/h$, comparable to the most massive cluster bin studied by \citet{wetzel13}.
In Fig.~\ref{fig:mstar} we show our subhalo stellar mass distributions: all three bins have similar distributions with mean
mass $\sim 10^{11} M_\odot$, with a very slight shift to larger stellar mass for larger $r_c$.
For this host mass and subhalo stellar mass, Fig.~10 of \citet{wetzel13} indicates that 55\% of quiescent satellites quenched as satellites (40\% in the current host and 15\% as satellites of a prior host; 45\% quenched while isolated).  This estimate appears to be consistent with our findings that a large fraction of redMaPPer satellites quenched prior to entering their current hosts.  It will be interesting, therefore, to perform the same measurement for satellites with lower stellar masses, where the \citeauthor{wetzel13} model predicts nearly $\sim 100\%$  quenched as satellites.  Such measurements should be possible with ongoing imaging surveys like DES, HSC or DECALS, or future surveys like LSST.  Within the range of satellite stellar masses in our data (see Figure 8), we have split the sample into two with median stellar masses of about $4.6\times 10^{10}$ and $1.7\times 10^{11}  M_\odot /h$. The results are shown in Fig.~\ref{fig:stellarmass_split}; it is evident that deeper imaging data that obtains cluster satellites with lower stellar mass is needed to explore possible trends. 

Another reason it will be interesting to pursue similar measurements for lower-mass galaxies is that  the \citet{wetzel13} model requires a delay between the time when star forming satellites fell into their first host and the time when their star formation began to exponentially decay.
Their Figure~8 shows that this delay time is $\sim 2.5$ Gyr for our subhalos.
Since our SED fits imply that $\tquench \sim 6$ Gyr, the \citet{wetzel13} model would predict that first infall occured $\sim 8.5$ Gyr ago (i.e.\ at $z>1$) in order for satellites to be host-quenched. We see no evidence for such high redshifts of infall. For satellites with $M_\star\sim 10^{11}M_\odot$, a large fraction are predicted to quench as centrals, so our measurements are not inconsistent with the model's predictions.  However, if we find similar results for satellites with $M_\star\sim 10^{10}M_\odot$, it will be difficult to reconcile that behavior with host quenching models like the \citeauthor{wetzel13} model in which nearly $100\%$ of objects quenched as satellites.
One possible solution to this problem is if a significant fraction of the red satellites quenched as satellites of previous (smaller) hosts, rather than their current hosts.  For the host-satellite properties that match our data, \citet{wetzel13} estimate that only $\sim 15\%$ of such galaxies quenched as satellites in hosts different than their current hosts, and it is unclear whether this small fraction would be sufficient to account for the correlations observed in our sample. We intend to address some of these open questions in a future study that includes the use of deprojected, 3D correlations of satellites for quantitative comparisons with theoretical models.

\section*{Acknowledgments}

We would like to thank Gary Bernstein, Daniel Gruen, Eric Baxter, Juliana Kwan, Mike Jarvis, Vinicius Miranda, Vishal Kasliwal, Tamas Norbert Varga, Ravi Sheth and Risa Wechsler for helpful discussions. 
YF and BJ are partially supported by the US Department of Energy grant DE-SC0007901.  ND thanks the Institute for Advanced Study for hospitality during the course of this work.


\label{lastpage}


\begin{thebibliography}{}
\bibitem[Abadi et al.(1999)]{abadi99} Abadi, M.~G., Moore, B., \& Bower, R.~G.\ 1999, \mnras, 308, 947
\bibitem[Adhikari et al.(2014)]{adhikari14} Adhikari, S., Dalal, N., \& Chamberlain, R.~T.\ 2014, \jcap, 11, 019
\bibitem[Balogh et al.(2000)]{balogh00} Balogh, M.~L., Navarro, J.~F., \& Morris, S.~L.\ 2000, \apj, 540, 113
\bibitem[Campbell et al.(2016)]{campbell16} Campbell, D., Padmanabhan, N., \& van den Bosch, F.~C.\ 2016, in SnowPAC 2016: The Galaxy-Halo Connection, http://www.physics.utah.edu/snowpac/talks/campbell.pdf
\bibitem[Chamberlain et al.(2015)]{chamberlain15} Chamberlain, R.~T., Dalal, N., Hearin, A., \& Ricker, P.\ 2015, \mnras, 451, 1496
\bibitem[Clampitt et al.(2016)]{clampitt16a} Clampitt, J., Miyatake, H., Jain, B., \& Takada, M.\ 2016, \mnras, 457, 2391
\bibitem[Clampitt \& Jain(2016)]{clampitt16b} Clampitt, J., \& Jain, B.\ 2016, \mnras, 457, 4135
\bibitem[Cohn(2012)]{cohn12} Cohn, J.~D.\ 2012, \mnras, 419, 1017 
\bibitem[Cohn \& White(2014)]{cohn14} Cohn, J.~D., \& White, M.\ 2014, \mnras, 440, 1712 
\bibitem[Dalal et al.(2008)]{dalal08} Dalal, N., White, M., Bond, J.~R., \& Shirokov, A.\ 2008, \apj, 687, 12
\bibitem[Dekel \& Birnboim(2006)]{dekel06} Dekel, A., \& Birnboim, Y.\ 2006, \mnras, 368, 2
\bibitem[Diemer \& Kravtsov(2014)]{diemer14} Diemer, B., \& Kravtsov, A.~V.\ 2014, \apj, 789, 1
\bibitem[Dietrich et al.~(2012)]{dietrich12} Dietrich, J.~P., Werner, N., Clowe, D., et al.\ 2012, \nat, 487, 202
\bibitem[Donahue et al.(2016)]{donahue16} Donahue, M., Ettori, S., Rasia, E., et al.\ 2016, arXiv:1601.04947
\bibitem[Erben et al.(2013)]{erben13} Erben, T., Hildebrandt, H., Miller, L., et al.\ 2013, \mnras, 433, 2545
\bibitem[Evans \& Bridle(2009)]{evans09} Evans, A.~K.~D., \& Bridle, S.\ 2009, \apj, 695, 1446
\bibitem[Farouki \& Shapiro(1981)]{farouki81} Farouki, R., \& Shapiro, S.~L.\ 1981, \apj, 243, 32 
\bibitem[Gao et al.(2004)]{gao04} Gao, L., White, S.~D.~M., Jenkins, A., Stoehr, F., \& Springel, V.\ 2004, \mnras, 355, 819
\bibitem[Gao et al.(2005)]{gao05} Gao, L., Springel, V., \& White, S.~D.~M.\ 2005, \mnras, 363, L66
\bibitem[Gunn \& Gott(1972)]{gunn72} Gunn, J.~E., \& Gott, J.~R., III 1972, \apj, 176, 1
\bibitem[Hayashi et al.(2003)]{hayashi03} Hayashi, E., Navarro, J.~F., Taylor, J.~E., Stadel, J., \& Quinn, T.\ 2003, \apj, 584, 541
\bibitem[Hearin \& Watson(2013)]{hearin13} Hearin, A.~P., \& Watson, D.~F.\ 2013, \mnras, 435, 1313
\bibitem[Hearin et al.(2014)]{hearin14} Hearin, A.~P., Watson, D.~F., Becker, M.~R., et al.\ 2014, \mnras, 444, 729
\bibitem[Jauzac et al.(2012)]{jauzac12} Jauzac, M., Jullo, E., Kneib, J.-P., et al.\ 2012, \mnras, 426, 3369
\bibitem[Johnston et al.(2007)]{johnston07} Johnston, D.~E., Sheldon, E.~S., Wechsler, R.~H., et al.\ 2007, arXiv:0709.1159
\bibitem[Kuijken et al.(2015)]{kuijken15} Kuijken, K., Heymans, C., Hildebrandt, H., et al.\ 2015, \mnras, 454, 3500
\bibitem[Larson et al.(1980)]{larson80} Larson, R.~B., Tinsley, B.~M., \& Caldwell, C.~N.\ 1980, \apj, 237, 692
\bibitem[Li et al.(2015)]{li15} Li, R., Shan, H., Kneib, J.-P., et al.\ 2015, arXiv:1507.01464
\bibitem[Mandelbaum et al.(2006)]{mandelbaum06} Mandelbaum, R., Seljak, U., Cool, R.~J., et al.\ 2006, \mnras, 372, 758
\bibitem[Mandelbaum et al.(2008)]{mandelbaum08} Mandelbaum, R., Seljak, U., \& Hirata, C.~M.\ 2008, \jcap, 8, 006
\bibitem[Mandelbaum et al.(2010)]{mandelbaum10} Mandelbaum, R., Seljak, U., Baldauf, T., \& Smith, R.~E.\ 2010, \mnras, 405, 2078
\bibitem[Mandelbaum et al.(2016)]{mandelbaum16} Mandelbaum, R., Wang, W., Zu, Y., et al.\ 2016, \mnras, 457, 3200
\bibitem[Maraston et al.(2010)]{maraston10} Maraston, C., Pforr, J., Renzini, A., et al.\ 2010, \mnras, 407, 830
\bibitem[Melchior et al.(2015)]{melchior15} Melchior, P., Suchyta, E., Huff, E., et al.\ 2015, \mnras, 449, 2219
\bibitem[Miyatake et al.(2016)]{miyatake16} Miyatake, H., More, S., Takada, M., et al.\ 2016, Physical Review Letters, 116, 041301
\bibitem[Miyazaki et al.(2015)]{miyazaki15} Miyazaki, S., Oguri, M., Hamana, T., et al.\ 2015, \apj, 807, 22
\bibitem[Moore et al.(1996)]{moore96} Moore, B., Katz, N., Lake, G., Dressler, A., \& Oemler, A.\ 1996, \nat, 379, 613 
\bibitem[More et al.(2016)]{more16} More, S., Miyatake, H., Takada, M., et al.\ 2016, arXiv:1601.06063
\bibitem[Moustakas et al.(2013)]{moustakas13} Moustakas, J., Coil, A.~L., Aird, J., et al.\ 2013, \apj, 767, 50
\bibitem[Navarro et al.~(1997)]{nfw1997} Navarro, J.~F., Frenk, C.~S., \& White, S.~D.~M.\ 1997, \apj, 490, 493
\bibitem[Norberg et al.(2009)]{norberg09} Norberg, P., Baugh, C.~M., Gazta{\~n}aga, E., \& Croton, D.~J.\ 2009, \mnras, 396, 19
\bibitem[Nord et al.(2015)]{nord15} Nord, B., Buckley-Geer, E., Lin, H., et al.\ 2015, arXiv:1512.03062
\bibitem[Oguri et al.(2010)]{oguri10} Oguri, M., Takada, M., Okabe, N., \& Smith, G.~P.\ 2010, \mnras, 405, 2215
\bibitem[Pastor Mira et al.(2011)]{pastor11} Pastor Mira, E., Hilbert, S., Hartlap, J., \& Schneider, P.\ 2011, \aap, 531, A169
\bibitem[Rozo et al.(2009)]{rozo09} Rozo, E., Rykoff, E.~S., Evrard, A., et al.\ 2009, \apj, 699, 768
\bibitem[Rozo \& Rykoff(2014)]{rozo14} Rozo, E., \& Rykoff, E.~S.\ 2014, \apj, 783, 80
\bibitem[Rozo et al.(2015a)]{rozo15a} Rozo, E., Rykoff, E.~S., Becker, M., Reddick, R.~M., \& Wechsler, R.~H.\ 2015a, \mnras, 453, 38
\bibitem[Rozo et al.(2015b)]{rozo15b} Rozo, E., Rykoff, E.~S., Abate, A., et al.\ 2015b, arXiv:1507.05460
\bibitem[Rykoff et al.(2014)]{rykoff14} Rykoff, E.~S., Rozo, E., Busha, M.~T., et al.\ 2014, \apj, 785, 104
\bibitem[Rykoff et al.(2016)]{rykoff16} Rykoff, E.~S., Rozo, E., Hollowood, D., et al.\ 2016, arXiv:1601.00621
\bibitem[Salpeter(1955)]{salpeter55} Salpeter, E.~E.\ 1955, \apj, 121, 161
\bibitem[Sheldon et al.(2009)]{sheldon09} Sheldon, E.~S., Johnston, D.~E., Masjedi, M., et al.\ 2009, \apj, 703, 2232
\bibitem[Sheth \& Tormen(2004)]{sheth04} Sheth, R.~K., \& Tormen, G.\ 2004, \mnras, 350, 1385
\bibitem[Sif{\'o}n et al.(2015)]{sifon15} Sif{\'o}n, C., Cacciato, M., Hoekstra, H., et al.\ 2015, \mnras, 454, 3938
\bibitem[Simet et al.(2016)]{simet16} Simet, M., McClintock, T., Mandelbaum, R., et al.\ 2016, arXiv:1603.06953 
\bibitem[Viola et al.(2015)]{viola15} Viola, M., Cacciato, M., Brouwer, M., et al.\ 2015, \mnras, 452, 3529
\bibitem[Watson et al.(2015)]{watson15} Watson, D.~F., Hearin, A.~P., Berlind, A.~A., et al.\ 2015, \mnras, 446, 651
\bibitem[Wechsler et al.(2006)]{wechsler06} Wechsler, R.~H., Zentner, A.~R., Bullock, J.~S., Kravtsov, A.~V., \& Allgood, B.\ 2006, \apj, 652, 71
\bibitem[Wetzel et al.(2013)]{wetzel13} Wetzel, A.~R., Tinker, J.~L., Conroy, C., \& van den Bosch, F.~C.\ 2013, \mnras, 432, 336
\bibitem[Zhang et al.(2013)]{zhang13} Zhang, Y., Dietrich, J.~P., McKay, T.~A., Sheldon, E.~S., \& Nguyen, A.~T.~Q.\ 2013, \apj, 773, 115
\bibitem[Zu \& Mandelbaum(2015a)]{zu15a} Zu, Y., \& Mandelbaum, R.\ 2015, arXiv:1509.06758
\bibitem[Zu \& Mandelbaum(2015b)]{zu15b} Zu, Y., \& Mandelbaum, R.\ 2015, \mnras, 454, 1161
\end{thebibliography}
\end{document}